\documentclass[]{aa} 
\usepackage{graphicx}
\usepackage{txfonts}
\usepackage{natbib}
\usepackage{verbatim}
\usepackage{longtable}
\usepackage{lscape}
\usepackage{afterpage}
\usepackage{multirow}
\usepackage{verbatim}
\bibpunct{(}{)}{;}{a}{}{,}

\begin{document}

\title{Analysis of the emission of very small dust particles from Spitzer spectro-imagery data using 
Blind Signal Separation methods\thanks{This work is based on observations made with the Spitzer Space
Telescope, which is operated by the Jet Propulsion Laboratory, California Institute of Technology 
under a contract with NASA.}}

\author{O. Bern\'e \inst{1}
	\and 
	C. Joblin\inst{1}
	\and
	Y. Deville\inst{2}
	\and 
	J. D. Smith\inst{3}
	\and
	M. Rapacioli\inst{4}
	\and
	J. P. Bernard\inst{1}
	\and 
	J. Thomas\inst{2}
	\and
	W. Reach\inst{5}
	\and
	A. Abergel\inst{6}}

\offprints{\\ O.~Berne, \email{olivier.berne@cesr.fr}}
\institute{
Centre d'Etude Spatiale des Rayonnements, CNRS et Universit\'e Paul 
Sabatier Toulouse~3, Observatoire Midi-Pyr\'en\'ees, 9 Av. du Colonel Roche, 
31028 Toulouse cedex 04, France
\and
Laboratoire d'Astrophysique de Toulouse-Tarbes, CNRS et Universit\'e Paul 
Sabatier Toulouse~3, Observatoire Midi-Pyr\'en\'ees, 14 Av. Edouard Belin,
31400 Toulouse, France
\and
Steward Observatory, University of Arizona, Tucson, AZ 85721
\and
Technische Universit\"at Dresden, Institut f\"ur Physikalische Chemie und Elektrochemie, 
01062 Dresden, Germany
\and
Spitzer Science Center, MS 220-6, California Institue of Technology, Pasedena, CA 91125
\and
Institut d'Astrophysique Spatiale, CNRS et Universit\'e de Paris Sud, B\^at 109, 91405 Orsay Cedex, France
}

\date{Received ?; accepted ?}

\abstract
{This work was conducted as part of the \emph{SPECPDR} program, dedicated to the study of 
very small particles and astrochemistry, in Photo-Dissociation Regions (PDRs).}
{We present the analysis of the mid-IR spectro-imagery observations of Ced 201, NCG 7023 
East and North-West and $\rho$ Ophiuchi West filament.}
{Using the data from all four modules of the InfraRed Spectrograph onboard the 
\emph{Spitzer Space Telescope}, we produced a spectral cube ranging from 5 to 35 $\mu$m, 
for each one of the observed PDRs. The resulting cubes were analysed using 
\emph{Blind Signal Separation} methods (NMF and \emph{FastICA}).}
{For Ced 201, $\rho$ Ophiuchi West filament and NGC 7023 East, we find that two signals can be 
extracted from the original data cubes, which are 5 to 35 $\mu$m spectra. The main features of 
the first spectrum are a strong continuum emission at long wavelengths, and a broad 7.8 $\mu$m 
band. On the contrary, the second spectrum exhibits the classical Aromatic Infrared Bands (AIBs) 
and no continuum. The reconstructed spatial distribution maps show that the latter spectrum is 
mainly present at the cloud surface, close to the star whereas the first one is located slightly deeper
 inside the PDR. The study of the spectral energy distribution of Ced 201 up to 100 $\mu$m suggests 
 that, in cool PDRs, the 5-25 $\mu$m continuum is carried by Very Small Grains (VSGs).
The AIB spectra in the observed objects can be interpreted as the contribution of neutral and 
positively-charged Polycyclic Aromatic Hydrocarbons (PAHs).}
{We extracted the 5 to 25\,$\mu$m emission spectrum of  VSGs in cool PDRs, these grains being 
most likely carbonaceous. We show that the variations of the mid-IR (5-35 $\mu$m) spectra of PDRs
can be explained by the photo-chemical processing of these VSGs and PAHs, VSGs being the 
progenitors of free PAHs.
}


\authorrunning{Bern\'e et al.}
\titlerunning{Mid-IR spectra of very small dust particles, blind signal separation methods
 applied to \emph{Spitzer} data}

\maketitle

\section{Introduction} \label{introduction}
The Aromatic Infrared Bands (AIBs), observed in emission in UV irradiated interstellar matter, 
have been attributed to Polycyclic Aromatic Hydrocarbons (PAHs) by 
\citet{leg84} and \citet{all85}. Later observations with the Infrared Space Observatory (ISO) 
\citep{kes96} have largely contributed to the study of these macromolecules and their evolution 
in the ISM, together with many fundamental studies involving both laboratory experiments
and quantum chemistry calculations. Although these studies
seem to confirm that the AIB carriers are PAH-like species, a precise identification
of the mixture of the interstellar molecules has not been possible yet.
In parallel to this quest for identification, \citet{ces00} gave strong evidence
of  the existence of Very Small carbonaceous Grains (VSGs),
from the ISO observations of the reflection nebula Ced 201. They proposed that these grains, emitting
broad bands and a continuum, are destroyed by UV radiation and/or shock waves freeing the AIB carriers
in the process.
In addition, they suggested that these grains could be present
everywhere in the ISM, but detectable in the mid-IR only under special conditions. Photo-Dissociation
Regions (PDRs) constitute the transition regions between UV-rich regions and molecular 
clouds. In PDRs with limited excitation, the infrared emission is dominated by dust particles. To emit 
at mid-IR wavelengths, these particles must be heated to high temperatures implying that they are very small
(from $\sim$ 1 to 10 nm) and not in thermal equilibrium with the incident radiation field.
PDRs are extremely useful to probe the properties of very small dust particles, because 
the local UV radiation field varies as a function of the distance from the star.
This change in the excitation conditions is expected to imply changes in the dust processing and therefore
provoke significant variations in the observed infrared spectrum.
The analysis of the observed spectral variations can provide valuable clues on the physical and
chemical properties of the emitting very small particles.
The study of such regions with a high spectral and spatial resolution was first made possible with ISOCAM 
onboard the Infrared Space Observatory (ISO) \citep{ces96}. Most analyses of the ISOCAM spectral
cubes have been achieved with simple mathematical decompositions of the observed spectra in bands (the AIBs)
and continua \citep{abe03}. However, more
sophisticated methods were used elsewhere to describe the observed spectrum on each pixel as the combination 
of the emission of several chemical populations. This approach requires the use of methods such as
Single Value Decomposition (SVD is applied to the covariance matrix of the data, which is
 equivalent to Principal Component Analysis). 
Applying SVD to the observations of different PDRs, \citet{boi01} and then \citet{rap05} (RJB hereafter)
were able to extract the spectra of different types of very small dust particles: 
PAH-like species and VSGs.
Although pioneering, this work was restricted to two objects of the ISOCAM database,
because of the limited efficiency of the method.

In the field of signal processing, quite a large number of algorithms have been developed
to extract the original signals from observations which are mixtures of these signals. Such algorithms are
referred to as \emph{Blind Signal Separation} (BSS) methods. The most widely used
are Independent Component Analysis (ICA) approaches, with algorithms like \emph{FastICA}  \citep{hyv99}.
The use of ICA is so widespread that it is sometimes presented as the only class of BSS methods,
although some other algorithms like Non-negative Matrix Factorisation  (NMF) \citep{lee99} 
have proven to be efficient.

This paper presents the results of a study conducted with the newly available 
spectro-imagery data from the Infrared Spectrograph (IRS) \citep{hou04} onboard 
\emph{Spitzer}.
Our goal is to use both the spectral and spatial information available in IRS data cubes
to reveal the origin of the observed spectral variations.
The IRS spectrograph provides several improvements when compared to ISOCAM. 
These are a higher sensitivity (about 0.06 mJy from 6 to 15 $\mu$m, and 0.4 mJy from 15 to 38 $\mu$m),
a better spectral resolution ($\frac{\lambda}{\Delta~\lambda} \sim 60-127$ in the low resolution mode) and
 a range of wavelengths extended to 35 $\mu$m.
We use two classes of BSS algorithms, i.e. ICA and NMF, to extract the spectra of the different
populations of very small particles. This work, carried out on four PDRs, extends
previous studies on the following aspects:
\begin{enumerate}
\item[i)]
we have gained in sensibility and spectral resolution;
\item[ii)]
a wider range of wavelengths is available;
\item[iii)]
the BSS methods are more powerful than SVD-based methods;
\item[iv)]
as a consequence, our results involve fainter objects.  
\end{enumerate}

After describing the observations and data reduction in Sect. 2, we define the BSS methods
 we have used for our analysis in Sect. 3.
We then present the results we have obtained applying 
these methods to the data (Sects. 4-5) and provide evidence of their
efficiency (Sect. 6). We show that the extracted spectra can be attributed 
to different populations of dust (Sect. 7). Using the extracted
in the studied PDRs in Sect. 8.




\section{Observations} \label{observations}

\subsection{Selected PDRs} \label{pdrs}

This study was conducted  
as part of the \emph{SPECPDR}\footnote{http://www.cesr.fr/~joblin/SPECPDR\_public/Home} program \citep{job05}
in which a total of 11 PDRs were observed with all the instruments onboard \emph{Spitzer}. 
The PDRs selected  for this paper, Ced 201, the $\rho$-Ophiuchi filament, NGC 7023 
East and North,  have mild radiation 
fields (200 to 1300 in units of the \citet{hab68}, $G_0$=1.6 $\times$ 10$^{-10}$ W cm$^{-2}$)
 and relatively simple geometries. They were chosen among the 11 observations because their
 IR emissions show the strongest spectral changes
across the PDR, thus providing the best information on the nature of the carriers.

\subsubsection{Ced 201} \label{Ced201}

Ced 201 is a compact reflection nebula
lying in the Cepheus constellation about 420 pc from the Sun \citep{cas91}. 
Unlike common reflection nebulae, which give birth to their own star, Ced 201 is the result of a chance
encounter between a B9.5\,V  star (BD+69$^{\circ}$1231) and a molecular cloud, at a relative velocity 
of about 12 km~s$^{-1}$ \citep{wit87}.
Because of this particularity, Ced 201 is a very attractive object for the spectroscopic study of
interstellar dust. \citet{ces00} revealed the presence of very small
carbonaceous grains, with a spectrum exhibiting bands and continuum emission, identified as the 
general mid-IR VSG spectrum  by RJB.

\subsubsection{NGC 7023 East and North-West} \label{NGC 7023}

NGC 7023 is a well studied reflection nebula located at 430 pc from the sun \citep{vda97}. Three PDRs are 
visible around the Herbig Be illuminating star, HD 200775. The brightest one, NGC 7023 North-West
(NGC 7023-NW hereafter), lies about 40 '' North-West 
of the star and another one lies around 70 '' South. The last one, NGC 7023 East (NGC 7023-E hereafter), 
is located further East of the star ($\sim$170'').
The spectroscopic studies of the North-West and South PDRs were performed by
\citet{ces96} using ISOCAM, and by \citet{wer04} with IRS, leading to the discovery of new infrared features.
In the present paper, we analyse the IRS spectro-imagery observations of NGC 7023-E and NGC 7023-NW. 

\subsubsection{The $\rho$-Ophiuchi filament} \label{NGC 7023}

The $\rho$-Ophiuchi molecular cloud is a nearby star-forming region. The western part of this cloud
is bounded by a filament structure (Oph-fil hereafter), forming an edge-on PDR, illuminated by the
B2 HD147889 star \citep{abe99, hab03}.

\subsection{IRS observations and data reduction} \label{obsred}

The four PDRs were observed using IRS onboard \emph{Spitzer}, in the low resolution
 ($\frac{\lambda}{\Delta~\lambda}=60-127$) mode.
The data were obtained in the spectral mapping mode with a spatial resolution 
from 3.6 to 10.2 arcseconds), where the spacecraft moves 
the spectrograph's slit before each integration, in order to cover a given area of the extended PDR. Each full
spectral cube was assembled using the CUBISM software \citep{smi06}.
The original Basic Calibrated Data (BCD) files are from the S12 pipeline of the  \emph{Spitzer} Science Center.
Calibration was achieved using the FLUXCON tables provided by the \emph{Spitzer} Science Center
together with the retrieved data. Additional 
corrections for extended sources were applied \citep{smi04} and the remaining bad pixels
were removed by hand.
In the low spectral resolution mode, IRS is divided in two modules: 
the Short wavelengths Low resolution (SL) module, and Long wavelengths Low resolution (LL) module.
Each module is sub-divided into two orders: SL2 (5 to 8.7 $\mu$m) and SL1 (7.4 to 14.5 $\mu$m) for SL and
LL2 (14 to 21.3 $\mu$m) and LL1 (19.5 to 38 $\mu$m) for LL.
For Ced 201, both SL and LL modules were used, providing the data from 5 to 38 $\mu$m.
For NGC 7023-NW we assembled the SL data from the PID 28 \emph{Spitzer} program\footnote{\emph{Spectral Line
Diagnostics of Shocks and Photon-Dominated Regions}, PI Giovanni Fazio.} 
with the LL2 order data obtained as "bonus" in the SPECPDR program, while the LL1 order
of the IRS was observing NGC 7023-E.
For NGC 7023-E and Oph-fil, the SPECPDR observations were only performed with the LL module,
to complement the available and good-quality ISOCAM spectral cubes.
The full spectral cubes, ranging from 5 to 38 $\mu$m, were constructed assembling the cubes from 
each module, and reprojecting the data on a grid with the lowest resolution (i.e. the LL grid for Ced
201 and NGC 7023-NW, and the ISOCAM grid for NGC 7023-E and Oph-fil). The spectral data from 35 to 38 $\mu$m
 is disregarded because of its poorer quality. The resulting set of data consists of four spectral cubes with
wavelengths ranging from 5 to 35 $\mu$m. Each spectral cube is thus a 3-dimensional matrix
$C (p_x, p_y, \lambda )$, which defines the variations of the recorded data with respect to the
wavelength $\lambda $, for each considered position with coordinates $(p_x, p_y)$ in the cube.

\subsection{Zodiacal light} \label{zodi}

The contribution from zodiacal light to the observations can become significant when looking at 
relatively faint objects. This is the case of Ced 201. Fortunately, this nebula is situated in a 
region of the sky which is slightly contaminated by zodiacal light. We however achieved a background
correction for this object. This consists in subtracting the spectrum from a nearby region of dark
sky (including background emission and zodiacal light) to the observations.
The other objects show stronger emission and are thus less subject to relative contamination by zodiacal 
light (less than 10~\% at 20~$\mu$m). For these objects no correction was applied.

\subsection{MIPS-SED and IRAS data} \label{mips}

We have combined the Infrared-Red Astronomical Satellite (IRAS) photometry and Multi-Band Imaging Photometer (MIPS,
onboard \emph{Spitzer}) data with the IRS data to build the spectral energy distribution of Ced 201, between 
5 and 100 $\mu$m. The IRS spectrum of Ced 201 was obtained by averaging the full IRS cube
(see Sect.~\ref{obsred}). The IRAS data was retrieved from the archive. Because the IRAS beam is larger than
the whole region observed in the spectral mapping mode with IRS, it is in principle not possible to compare
the fluxes observed with the two instruments. Thus we have made the assumption that the emission of
Ced 201 is dominated by the region close to the illuminating star and covered by IRS observations.
Therefore the flux calibration of the IRAS points at 12, 25, 60 and 100 $\mu$m points was achieved by scaling
the 12 and 25 $\mu$m points to the IRS-LL data.
The MIPS-SED cubes were built stacking the BCD files from the S12 pipeline. The MIPS-SED intensity spectrum
was done by averaging the data over the common region with IRS.

\section{Blind Signal Separation Methods} \label{bss}

\subsection{Problem overview}\label{PbOverview}
\emph{Blind Signal Separation} is commonly used to restore a set of unknown "source" signals from
a set of observed signals which are mixtures of these source signals,
with unknown mixture parameters \citep{hyv01}.
It has e.g. been applied in acoustics for unmixing recordings, or in the
biomedical field for separating
mixed electromagnetic signals produced by the brain \citep{saj04} . BSS is most
often achieved using ICA methods.
In particular,
an algorithm called \emph{FastICA} \citep{hyv99} has proven to be efficient
for recovering source signals.
An alternative class of methods for achieving BSS
is NMF, which was introduced in \citet{lee99} and then extended by a few authors.

In the field of Astrophysics, ICA has been sucessfully used for spectra
discrimination in infrared spectro-imagery of Mars ices \citep{for05}, elimination
of artifacts in astronomical images \citep{fun03}
or extraction of cosmic microwave background signal in \emph{Planck} simulated data \citep{mai02}.
To our knowledge,
NMF has not yet been applied to astrophysical problems.
However,
it has been used to separate spectra in other application fields,
e.g. for magnetic resonance chemical shift imaging of the human brain
\citep{saj04}
or for analysing wheat grain spectra
\citep{gob05}

The simplest version of the BSS problem concerns so-called "linear
instantaneous" mixtures.
It is modelled as follows:
\begin{equation}
\label{general}
X=AS
\end{equation}
where $X$ is an $m\times n$ matrix containing $n$ samples
of $m$ observed signals, $A$ is an $m\times r$ mixing matrix
and $S$ is an $r \times n$ matrix
containing $n$ samples
of $r$ source signals.
The observed signal samples are considered to be linear combinations of the source signal samples (with the same sample index). 
It is assumed that
$r \leq m$
in most investigations, including this paper.
The objective of BSS algorithms is then to recover the source matrix $S$ and/or the mixing matrix $A$
from $X$.

The correspondence between the generic BSS data model
(\ref{general}) and the
3-dimensional spectral cube
$C (p_x, p_y, \lambda )$
to be analysed in the present paper may be defined as follows.
Each observed signal, consisting of all samples available for this signal,
corresponds to
a row of $X$ in Eq.
(\ref{general}).
In this paper, the sample index
is associated
to the wavelength
$\lambda$,
and each observed signal
consists of the overall spectrum
recorded for a cube pixel
$(p_x, p_y)$.
Moreover,
each observed spectrum is a linear combination of
"source spectra", 
which are respectively associated to each of the (unknown) types of
dust particles that contribute to the recorded spectral cube.
Therefore, the recorded spectra may here be expressed according to
(\ref{general}), with unknown
combination coefficients in
$A$,
unknown
source spectra in
$S$
and an unknown number
$r$
of source spectra.
Sections \ref{fastica} and \ref{thNMF} introduce the two BSS methods we have tested
on these data, i.e. FastICA and an NMF algorithm.

\subsection{FastICA}\label{fastica}
%
\emph{FastICA}
is a statistical BSS method intended for
stationary, non-Gaussian and mutually statistically independent
random signals \citep{hyv99}.
It is expressed for zero-mean signals hereafter. In practice,
such signals are obtained by first subtracting
the sample mean of each observed signal to all
signal samples in the corresponding
row of $X$ in
(\ref{general}).
For the sake of simplicity, the notation $X$ refers to these
zero-mean signals hereafter.

The next step of FastICA consists in applying SVD to the  covariance matrix of the observed data, i.e. in deriving
a matrix $Z$ of transformed signals defined as
\begin{equation}\label{icamodel}
Z = M X
\end{equation}
where $M$ is selected so that: i) all signals in
$Z$ are mutually uncorrelated, ii) each of these signals has unit power
and iii) the number of signals in $Z$ is $ \hat{r}$.
The value of $ \hat{r}$ is selected as follows.
When applied to the $m$ signals in $X$,
SVD intrinsically yields $m$ output components.
Keeping all these components therefore corresponds to selecting
$ \hat{r} = m$.
Instead,
if $ r < m$,
one may choose to only
keep the $ \hat{r}$ output components which have the
highest powers,
with $ \hat{r}$ selected so that
$ r \leq \hat{r} < m$
(see details on p. 129 of \citet{hyv01} ).
This reduces the dimensionality of the processed data and allows one to combine
the following two features:
i) using
$ \hat{r} \geq r$ still makes it possible to recover all source
signals from $Z$
and
ii)
selecting
$ \hat{r} < m$
decreases in $Z$
the influence of noise components which exist in real
recordings $X$ but were not taken into account in the above data model
(\ref{icamodel}).

The basic version of FastICA then extracts a first source signal from
the matrix
$Z$.
The criterion used to this end consists in maximising the non-Gaussianity
of an output signal
defined as a linear combination of the signals in
$Z$.
Therefore,
denoting $z$ a column of $Z$ corresponding to a given sample index,
the corresponding sample of the output signal reads
\begin{equation}
\label{eq-fastica-output-onesig-onesamp}
y = d^Tz,
\end{equation}
where the column vector
$d$ is constrained to have unit norm.
Several versions of the FastICA
method have been defined, depending on which
parameter is used to measure the non-Gaussianity of $y$.
The most standard parameter is the
absolute value of the non-normalised kurtosis, defined for
a zero-mean signal $y$ as
\begin{equation}
\label{kurt}
Kurt(y)=E[y^4] - 3 ( E[y^2] )^2
\end{equation}
where $E [.]$ stands for expectation. 
Various algorithms may then be used for
adapting $d$ so as to maximise that absolute kurtosis parameter.
Before the FastICA algorithm was introduced,
\citet{del95}
optimised it by using a standard gradient ascent procedure.
FastICA is an alternative,
fixed-point, optimisation algorithm
described in
\citet{hyv99}.
It has been shown to yield much faster and more reliable convergence than
gradient procedures. Moreover, it does not require one to select any tunable
parameter (such as the adaptation gain of gradient algorithms). 
Once a first source signal has thus been extracted as an output signal $y$, \emph{FastICA}
removes its contributions from all observed signals contained by $X$. This yields a matrix
$X^{\prime}$ which only contain $r-1$ sources.
The same procedure as above is then applied to $X^{\prime}$ in order to extract another source.
This "deflation" procedure is repeated until all sources are extracted from the observations.

\subsection{Non-Negative Matrix Factorisation}\label{thNMF}

Unlike ICA, NMF is based on the assumed non-negativeness of the source signals and
mixing coefficients without requiring independence of the source signals.
It aims at recovering the $r$ source signals by approximating 
the supposedly non-negative matrix $X$ with
the following factorisation:

\begin{equation}
\label{nmf1}
X\approx W H
\end{equation}
where $W$ is a $m \times r$ non-negative
weight
matrix and $H$ is a $r \times n$
non-negative matrix
of approximated "source" signals.
The approximation quality in
eq. (\ref{nmf1})
can be optimised by adapting the non-negative matrices
$W$ and $H$
so as to
minimise the
squared Euclidian distance $\|X-WH\|^{2}$
or the divergence $D(X|WH)$ \citep{lee01},
defined as

\begin{equation}
\label{def1}
\|X-WH\|^{2}=\sum_{ij}(X_{ij}-(WH)_{ij})^{2}
\end{equation}

and

\begin{equation}
\label{def2}
 D(X|WH)=\sum_{ij}(X_{ij}\log \frac{X_{ij}}{(WH)_{ij}}-X_{ij}+(WH)_{ij}).
\end{equation}

The adjustement of $W$ and $H$ can be achieved using classical gradient descent but
can be problematic (see Lee \& Seung 2001).
Therefore, the authors have proposed a new algorithm showing
that the Euclidian distance is non increasing under the
iterative update rule

\begin{equation}
\label{def3}
H_{a\mu} \leftarrow H_{a\mu} \frac{(W^{T}X)_{a\mu}}{(W^{T}WH)_{a\mu}}, W_{ia} \leftarrow W_{ia} \frac{(XH^{T})_{ia}}{(WHH^{T})_{ia}}
\end{equation}
and that the divergence is also non increasing under the rule

\begin{equation}
\label{def4}
H_{a\mu} \leftarrow H_{a\mu} \frac{\sum_{i}W_{ia}X_{i\mu}/(WH)_{i\mu}}{\sum_{k}W_{ka}},
W_{ia} \leftarrow W_{ia} \frac{\sum_{\mu}H_{a\mu}X_{i\mu}/(WH)_{i\mu}}{\sum_{\nu}H_{a\nu}} 
\end{equation}

Thus, the following iterative algorithm can be derived from this result in order to
 minimise either the euclidian distance or divergence:
\begin{enumerate}
\item{fix $r$}
\item{initialise randomly matrices $W$ and $H$}
\item{update these matrices with the update rule (\ref{def3}) or (\ref{def4})}
\item{if convergence is reached, then end. Otherwise go back to Step 3.}
\end{enumerate}
When convergence is reached, $H$ provides an approximation of "source" signals.

\section{Application of the BSS methods} \label{choice}

\subsection{Suitability of BSS methods for analysis of \emph{Spitzer}-IRS cubes}
\label{suit}

In order to apply the above BSS methods to the IRS data cubes, it is necessary to 
make sure that the "linear instantaneous" mixture condition is fulfilled 
(see Sect.~\ref{PbOverview}).
Here we consider that each observed spectrum is a linear combination of "source spectra", which are
due to the emission of different populations of dust particles.
The main effect that can disturb the linearity of the model is radiative transfer as shown by
\citet{nuz}, because of the non-linearity of the equations.
In our case however, this effect is not expected to be a major concern. 
The mid-IR emission of dust particles occurs in the external layers of the cloud i.e. at
 optical depth lower or comparable with H$_{2}$ emission \citep{job07}.
Using the Meudon PDR code (Le Bourlot et al. 1993, Le Petit et al. 2006) we found
that $H_{2}$ emission occurs at extinctions of $A_{V}\leq$~0.2 in NGC 7023-E and
Oph-fil. For Ced 201 which is a particular case, as the star has penetrated the cloud, we use the
visual extinction $A_V$~=~0.21 of the illuminating star BD+69$^{\circ}$1231  derived by \citet{wit87}.
From this we conclude that very small particles emit in the infrared at a maximum visual extinction of
$A_V\sim$~0.2 in the studied PDRs.
We can then roughly estimate $\tau_{9.7}$, the optical depth of the 9.7 $\mu$m silicate
band which is the main absorption in the mid IR range.
Typically, the $\tau_{9.7}/A_V$ ratio is between $\sim$~10 and $\sim$~20
\citep{mat90}, which yieds $\tau_{9.7}<$~2~$\times$~10~$^{-2}$. This value
 is very small, indicating that the effect of radiative transfer on the mid-IR 
spectra is very weak.
Thus, the source spectra are not significantly affected by this reabsorption, and we conclude 
that the "linear instantaneous" mixture model is fulfilled. This is validated 
\emph{a posteriori} as we show that the extraction
has the same efficiency around 10 $\mu$m, where the silicates absorb, as at the other wavelengths 
(see Fig.\ref{residu}).

\subsection{Preliminary tests}\label{test}

To test the BSS methods presented in Sect.~\ref{bss}, we generated
 two synthetic spectra:
one representing a PAH-type emission spectrum, consisting of Lorentzian profiles at 6.2~$\mu$m, 
7.7~$\mu$m, 8.6~$\mu$m and 11.3~$\mu$m, and one representing the emission of VSGs (RJB),
consisting of a broad emission band at 7.8~$\mu$m, the 11.3~$\mu$m band and a second-order polynomial continuum. 
These spectra were mixed with a random $100 \times 2$ mixing matrix $A$, 
thus providing 100 artificial observed spectra, which are linear combinations of the initial
spectra. These spectra were then analysed using \emph{FastICA} and NMF.
With both methods, we managed to recover the original spectra efficiently.
In this case we have artificially mixed the sources, and therefore we can compare the extracted spectra
with these sources to measure the efficiency of the methods. 
With both methods, the correlation between the original and unmixed signals is excellent (correlation 
coefficients higher than 0.995).
However a realistic test must include the effect of noise. 
The noise in the IRS instrument is quite complex, and strongly depends on wavelength because of the 
different modules that are used. In order to simply quantify the effect of noise on the BSS we applied a white
spatially homogeneous noise. Though this is not representative of the real IRS noise wich would require a detailed 
study to be estimated, this provides good knowledge on the response of the algorithms to noise.
With both methods the results are still good for Signal to Noise Ratios (SNRs) over 2 (i.e. 3 dB).
As a comparison, the estimated IRS instrumental noise using the SpecPET tool\footnote{http://ssc.spitzer.caltech.edu/tools/specpet/} developed by the \emph{Spitzer} Science Center, is between 20 and 100 (13-20 dB) depending on the wavelength for Ced 201.
We have also estimated the noise using the spectrum of Ced 201 
presented on Fig.~\ref{recons} (position 2) and found a SNR of about 30 (14.8 dB) in the LL region which is consistent
 with the SpecPET estimation.
Below the limit SNR equal to 2, the efficiency of FastICA drops dramatically 
while NMF is still able to significantly recover the original signals (correlation coefficient 
greater than 0.85).

\subsection{Application to \emph{Spitzer} data}

We have applied \emph{FastICA} and NMF to the \emph{Spitzer} data.
The spectra from a given cube of observations $C(p_x,p_y,\lambda)$ are placed in the rows
of the $X$ matrix defined in Eq.~(\ref{general}) or (\ref{nmf1}).
To maintain the efficiency of the methods the noisiest spectra were removed.
Thus, $X$ contains the $m$ IRS spectra of a considered PDR, over $n$ points in wavelength.
For NMF, Euclidian distance or divergence were minimised using Lee \& Seung's
algorithm \citep{lee01} implemented with Matlab, to find $W$ and $H$.
The best results were found using the divergence criterion after about 1000 iterations
(which takes less than one minute with a a 3.2 GHz processor).
The value of $r$ (number of "source" spectra) is not imposed by the NMF method.
The strategy for extracting the sources was the following:
\begin{itemize}

\item{Apply the algorithm with a minimum number of assumed sources $ \hat{r} = 2$, to a given dataset,
providing 2 sources.}

$\rightarrow$ If the found solutions are physically coherent and
linearly independent we consider that at least $\hat{r}=2$ sources can be extracted.

$\rightarrow$ Else, we consider that the algorithm is not suited for analysis
(never occurred in our case).

\item{try the algorithm on the same dataset but with $\hat{r}=3$ sources.}

$\rightarrow$ If the found solutions are physically coherent and
linearly independent we consider that at least $\hat{r}=3$ sources can be extracted.

$\rightarrow$ Else, we consider that
only two sources can be extracted, extraction was over at $\hat{r}=2$ and thus $r=2$.

\item{ same as previous step but with $\hat{r}=4$ sources.}

$\rightarrow$ If the found solutions are physically coherent and
linearly independent we consider that at least $\hat{r}=4$ sources can be extracted.

$\rightarrow$  Else we consider that 
only three sources can be extracted, extraction was over at $\hat{r}=3$ and thus $r=3$.

$\ldots$
\end{itemize}

Physically uncoherent spectra exhibit sparse peaks (spikes) 
which cannot be PDR gas lines.
We found $r=3$ for NGC 7023-NW and $r=2$ for the other PDRs, implying that we could
respectively extract 3 and 2 spectra from these data cubes.
Similarly, with \emph{FastICA}, all sources were extracted with $r=2$ for Ced 201, Oph-fil 
and NGC 7023 E, and  $r=3$ for NGC 7023-NW.
Thus we were able to extract the "source" signals from the original cubes using both methods.
The choice of the method was therefore based on its suitability, considering the information
we have on the sources. In the following section, we argue on why NMF is more consistent in our case.

\subsection{Choice of the BSS method}\label{choice}

NMF and \emph{FastICA} have different constraints.
The strength of \emph{FastICA} is that the algorithm is guaranteed to converge towards a solution which yields 
separated spectra. However, this property only holds if the source signals are statistically independent,
due to the non-gaussianity criterion optimised by \emph{FastICA}.
This criterion is not ideal in our application, where the  "source" spectra
are likely to be correlated due to chemical similarities between the different emitting populations
(RJB). Moreover, the presence of noise can degrade the efficiency of the separation
with \emph{FastICA} (see Sect.~\ref{test}).
On the contrary,  the non-negativity of the "sources" and mixing coefficients, used in NMF, 
are consistent with the non-negativity of emission spectra and of their linear combination.
Thus NMF is likely to be more appropriate for the analysis
of our data cubes.
The only identified drawback of NMF is that its convergence point may depend on the random initialisation of the matrices (see Sect.~\ref{thNMF}). 
To test this effect, we have ran NMF on the data of NGC 7023-E using 200 different initialisation conditions.
We found that the outputs only slightly vary with the initialisation conditions.
Since NMF is most likely to have the greatest efficiency for our application, we now detail the results obtained with this
algorithm when applied to our data cubes.

\section{Results}
\label{Xtract}


In this section we present the extracted spectra for Ced 201, Oph-fil,  NGC 7023-E and 7023-NW 
(Figs.~\ref{ced201}-~\ref{ngc7023nw}) and their distribution maps.
These spectra are normalised to 1. This has to be done because they have a different
intensity depending on the considered pixel.
 The molecular hydrogen lines have been substracted in order to provide the spectra of dust only  and
 the hydrogen/dust interaction will be the subject of a subsequent paper \citep{job07}. 

\subsection{Extracted spectra}


\begin{figure}
\includegraphics[width=\hsize]{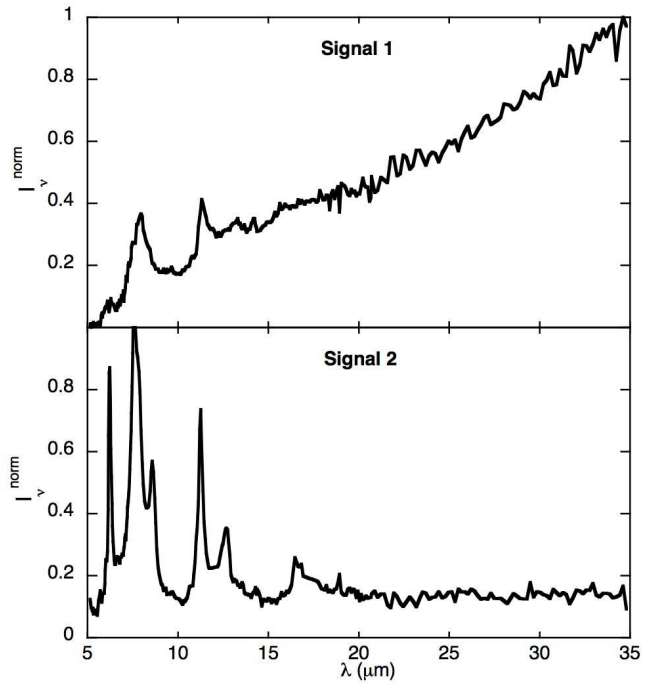}
\vspace{-0.0cm}
\caption{BSS extracted spectra in Ced 201.}
\label{ced201}
\vspace{-0.0cm}
\end{figure}

\begin{figure}
\includegraphics[width=\hsize]{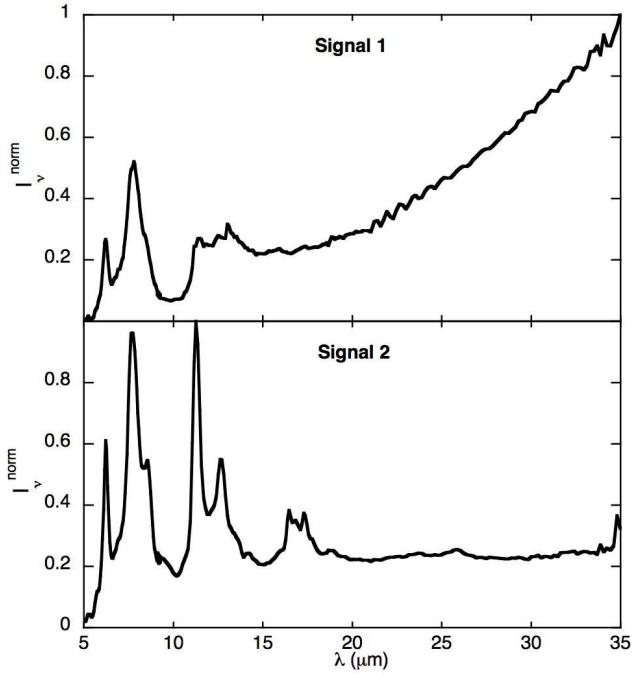}
\vspace{-0.0cm}
\caption{BSS extracted spectra in Oph-fil.}
\label{Oph-fil}
\vspace{-0.0cm}
\end{figure}

\begin{figure}
\includegraphics[width=\hsize]{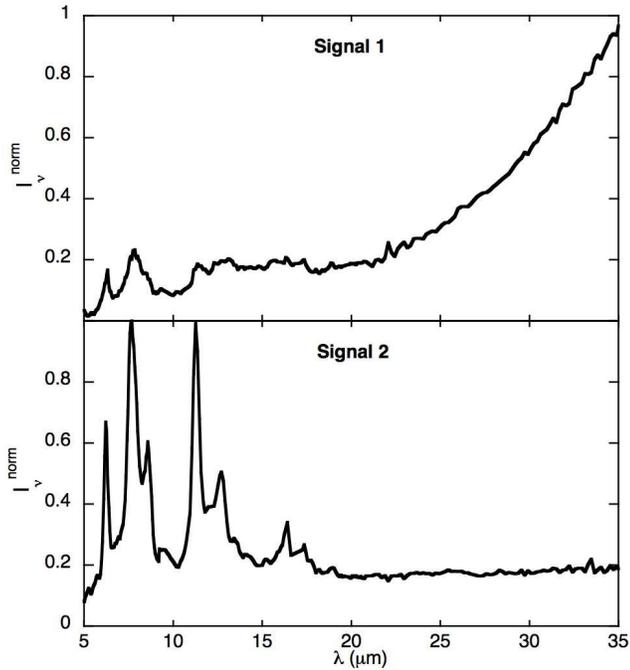}
\vspace{-0.0cm}
\caption{BSS extracted spectra in NGC 7023-E.}
\label{ngc7023}
\vspace{-0.0cm}
\end{figure}

\begin{figure}
\includegraphics[width=\hsize]{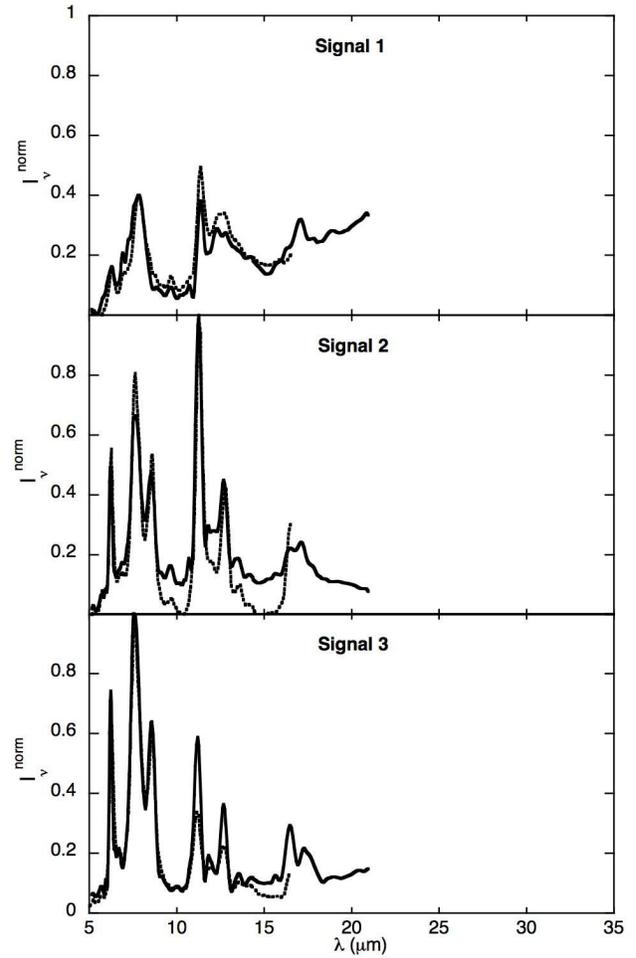}
\vspace{-0.0cm}
\caption{BSS extracted spectra in NGC 7023-NW. In this case, the 7.7 $\mu$m band of signal 1 was artificially normalised to
0.4 to be comparable with the spectra extracted in the other objects. The dotted line spectra are those extracted from the ISOCAM observations of the over-all NGC 7023 PDR by RJB. }
\label{ngc7023nw}
\vspace{-0.0cm}
\end{figure}

From each of the cubes of Ced 201, Oph-fil and NGC 7023-E, we were able to extract two spectra.
These independently found pairs of spectra have the same characteristics from one object to another. Moreover,
for each object, the two spectra exhibit different features.
The \emph{Signal 1} spectra, presented in 
 the upper part of Figs.~\ref{ced201}-\ref{ngc7023}
, are characterised by a rising continuum combined with broad  emission bands located at  6.2, 7.8, and 11.4 $\mu$m
(see Table~\ref{table2}).
On the contrary, the \emph{Signal 2} spectra, presented on the lower parts of Figs.~\ref{ced201}-\ref{ngc7023},
are dominated by the AIB emission at 6.2, 7.6, 8.6, 11.3, and 12.7 $\mu$m.
In NGC 7023-NW we could extract three signals from the original data.
These spectra are presented in Fig.~\ref{ngc7023nw}.
\emph{Signal 1} is very similar to \emph{Signal 1} found in the other PDRs studied above: it exhibits a broad 
7.8  $\mu$m band and a steep continuum.
\emph{Signal 2} and \emph{Signal 3} spectra are dominated by AIBs but with different relative intensities.
\emph{Signal 2} and \emph{Signal 3} are respectively dominated by the 11.3 and 7.6 $\mu$m bands.

\subsection{Spatial distributions of the extracted spectra}\label{distribution}

\begin{figure}
\begin{center}
\includegraphics[width=6cm]{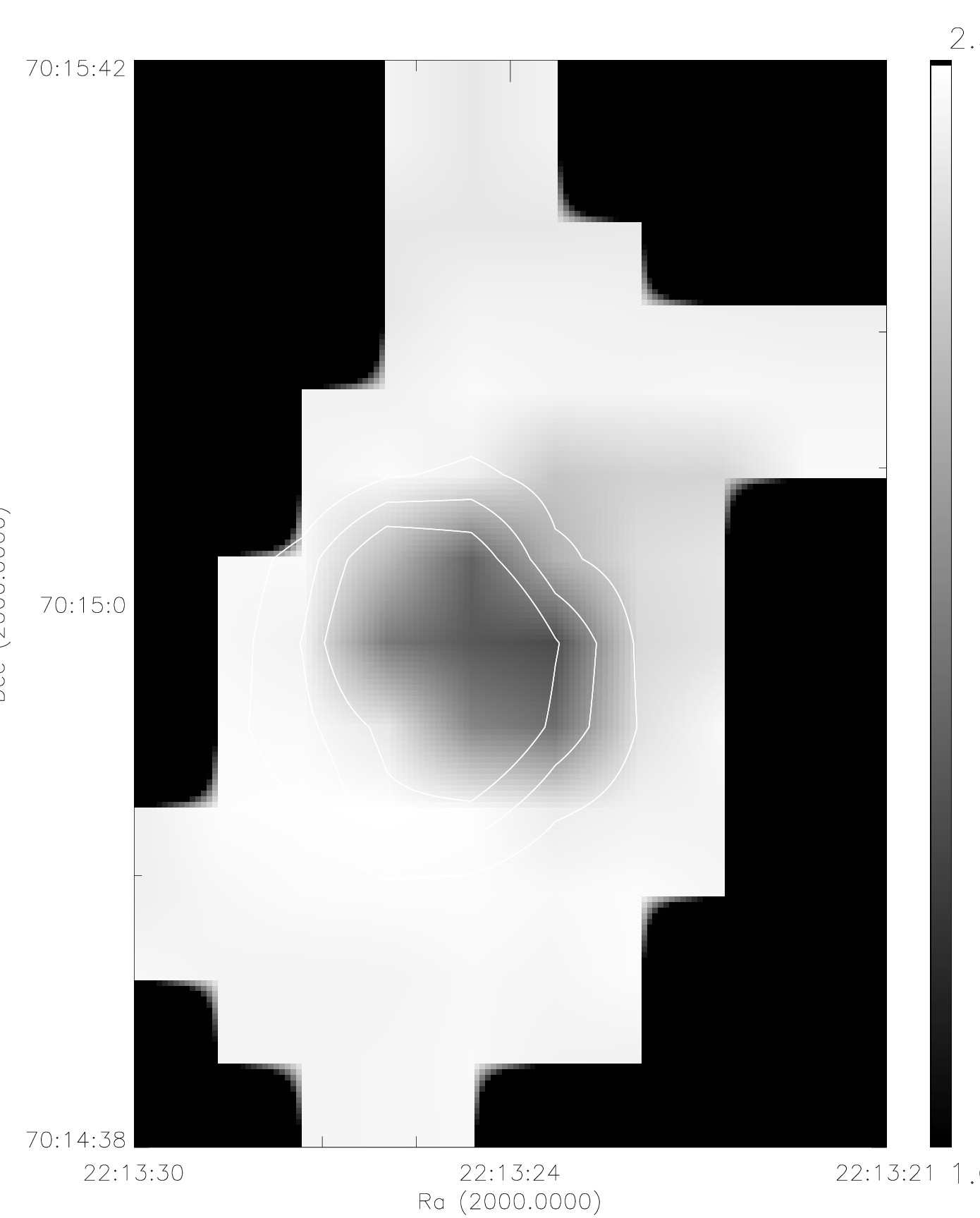}
\includegraphics[width=6cm]{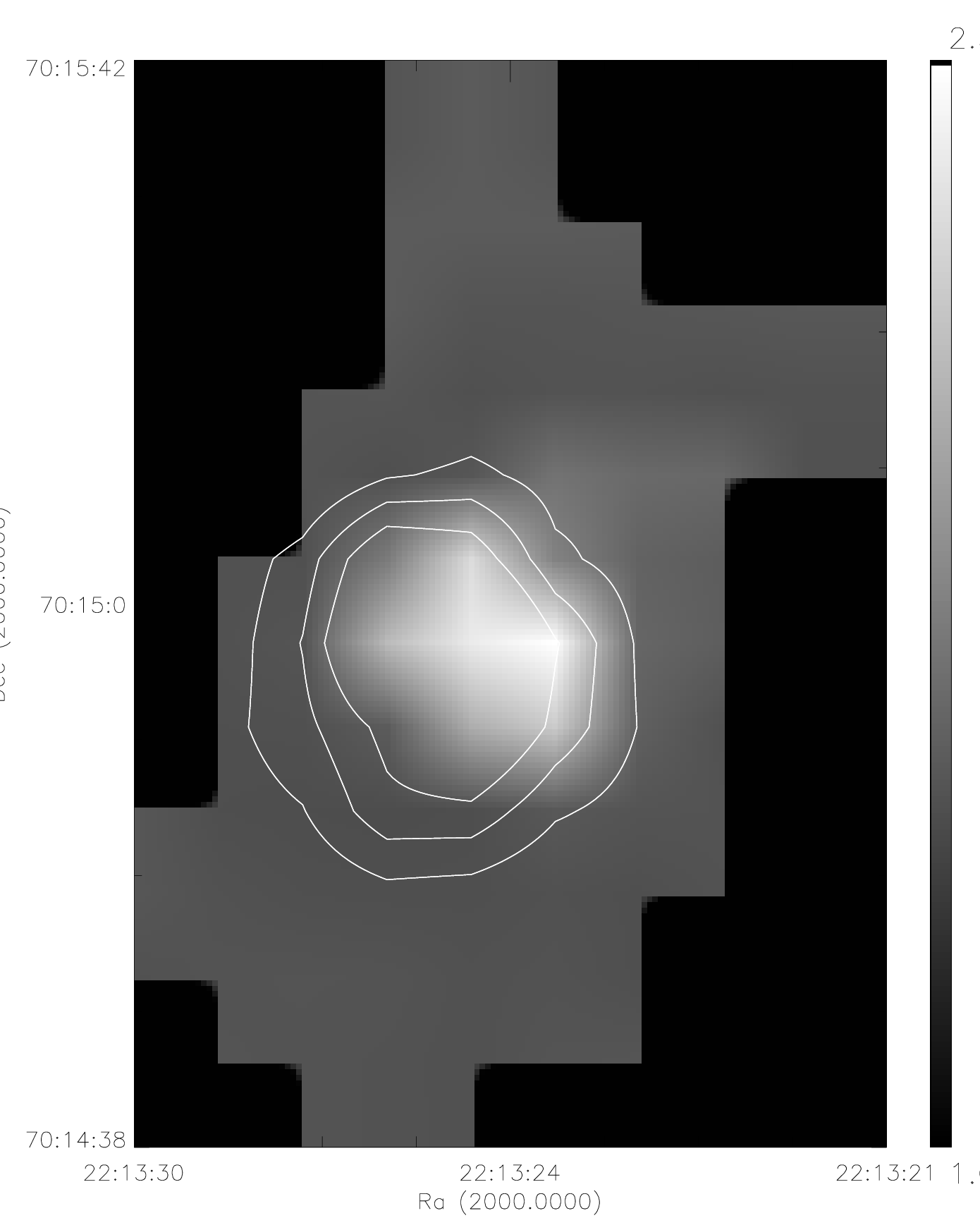}
\caption{Map of the emission ratios \emph{Signal 1}/\emph{Signal 2} (upper) and  \emph{Signal 2}/\emph{Signal 1} (lower) in Ced 201. 
The 5-35 $\mu$m integrated mid-IR emission is in contours (contours 1, 2, 3 are respectively 1, 1.5, 2 $\times$10$^{-4}$ W m$^{-2}$ sr$^{-1}$).
 The illuminating star is located at the center.
}
\label{map201}
\end{center}
\end{figure}

\begin{figure}
\begin{center}
\vspace{-1cm}

\includegraphics[width=6cm]{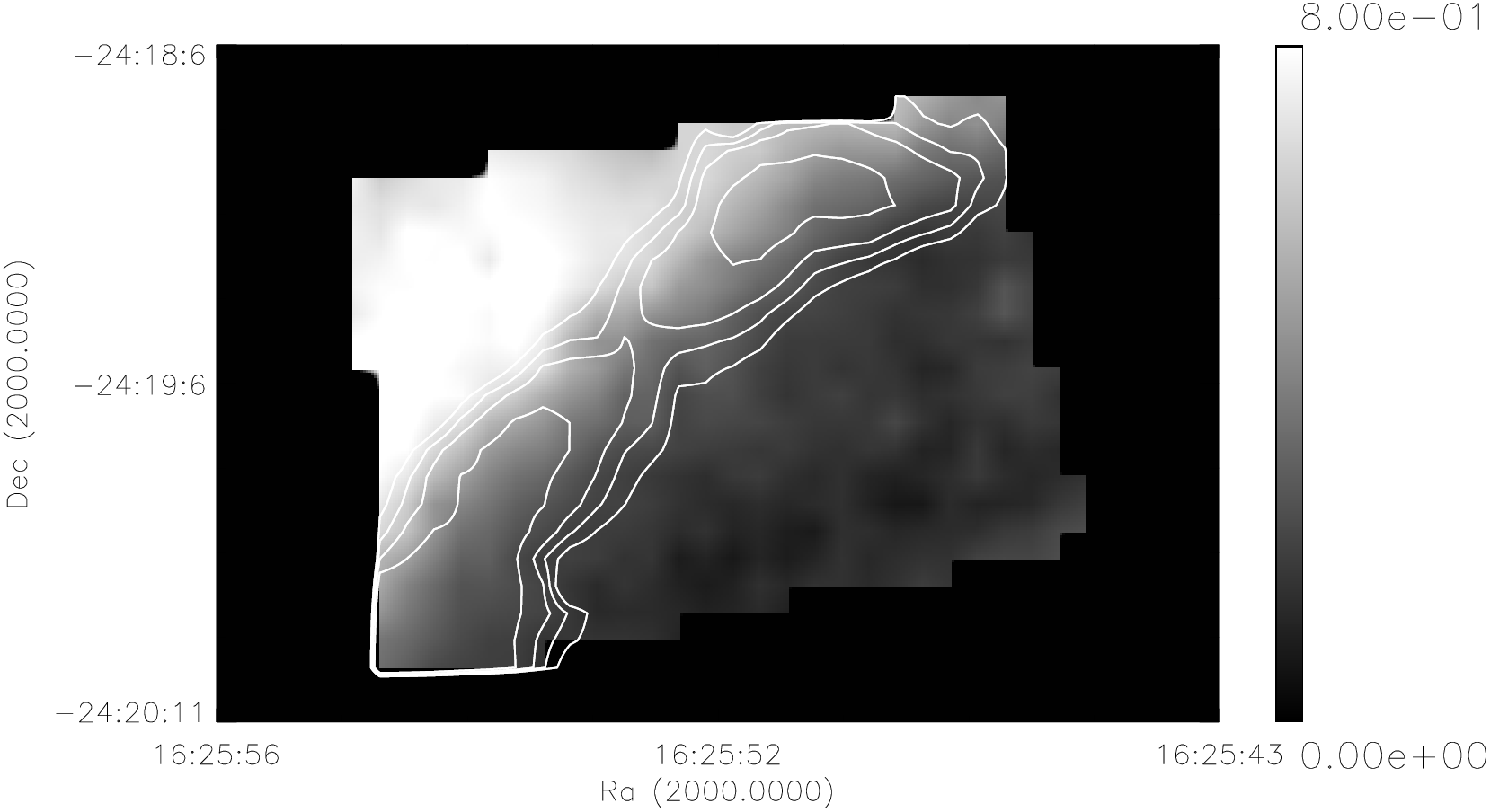}

\vspace{-3cm}
\includegraphics[width=6cm]{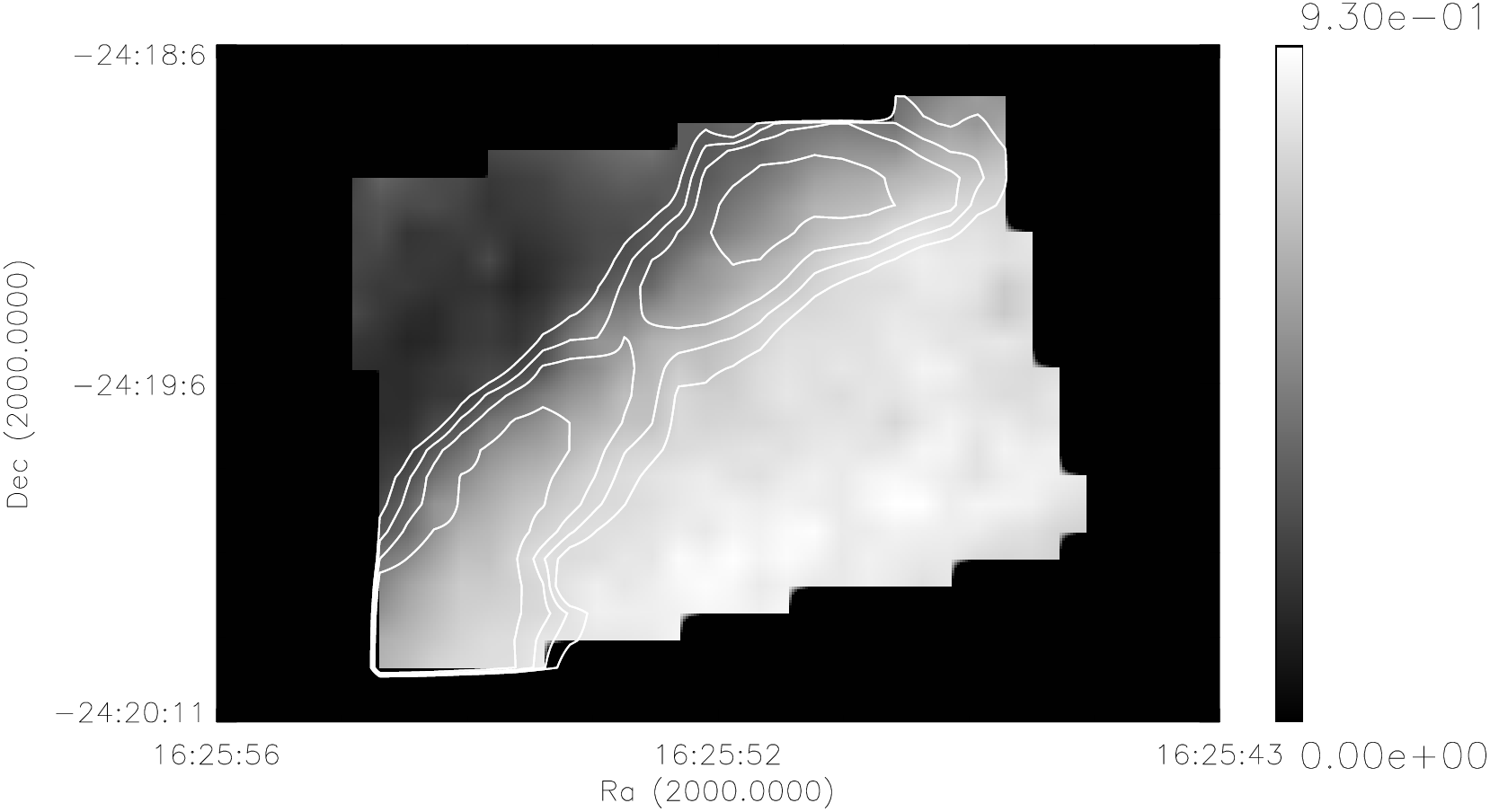}
\vspace{-0cm}
\vspace{-1cm}
\caption{ Map of the emission ratios \emph{Signal 1}/\emph{Signal 2} (upper) and  \emph{Signal 2}/\emph{Signal 1} (lower) in Oph-fil.
The 5-35 $\mu$m integrated mid-IR emission of the filament is in contours (contours 1, 2, 3, 4 are respectively
3.5, 3.7, 3.9, 4.2 $\times$10$^{-4}$ W m$^{-2}$ sr$^{-1}$).
The illuminating star lies South-West of the filament.}

\label{mapOph-fil}

\end{center}
\end{figure}

\begin{figure}
\begin{center}
\includegraphics[width=6cm]{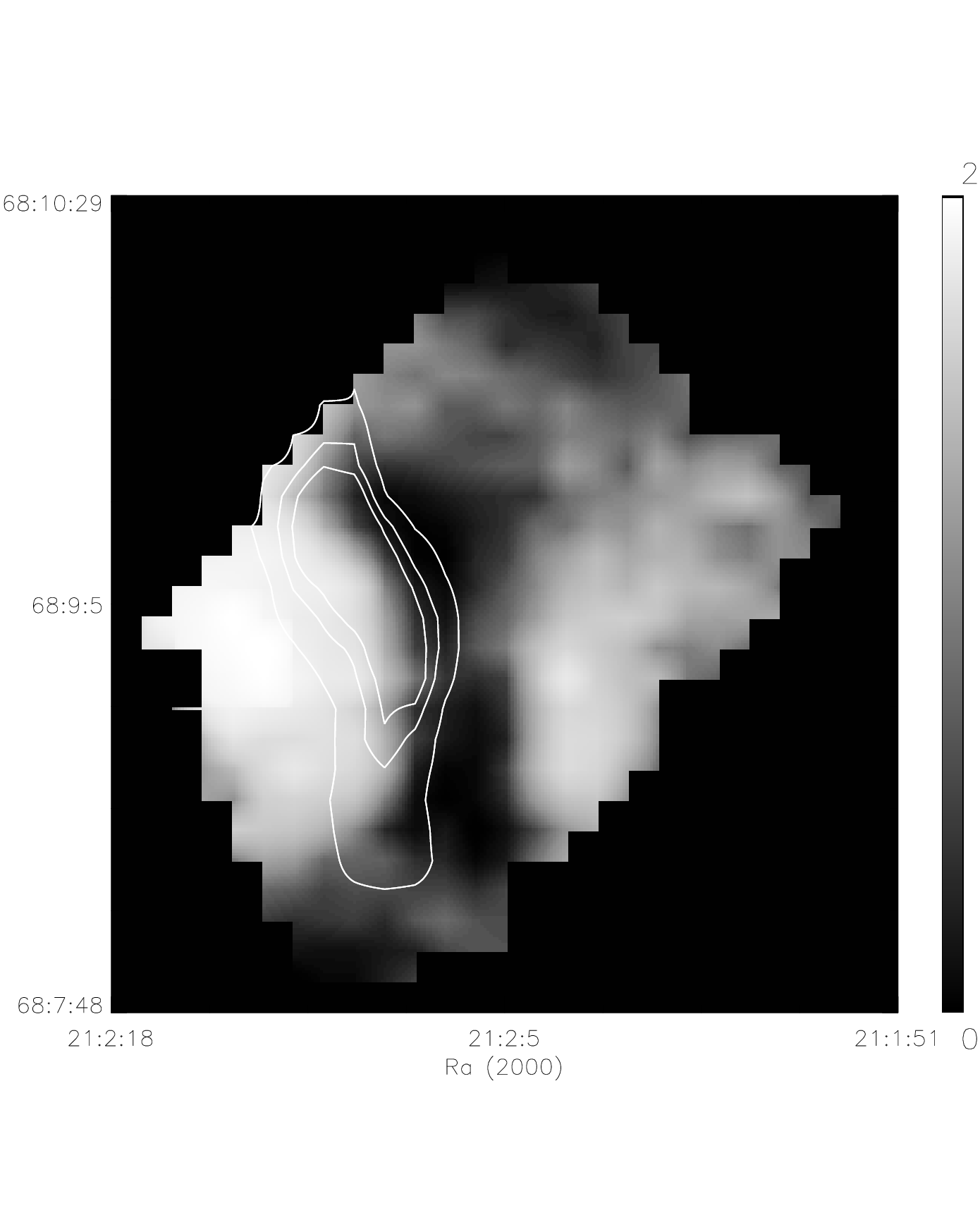}
\includegraphics[width=6cm]{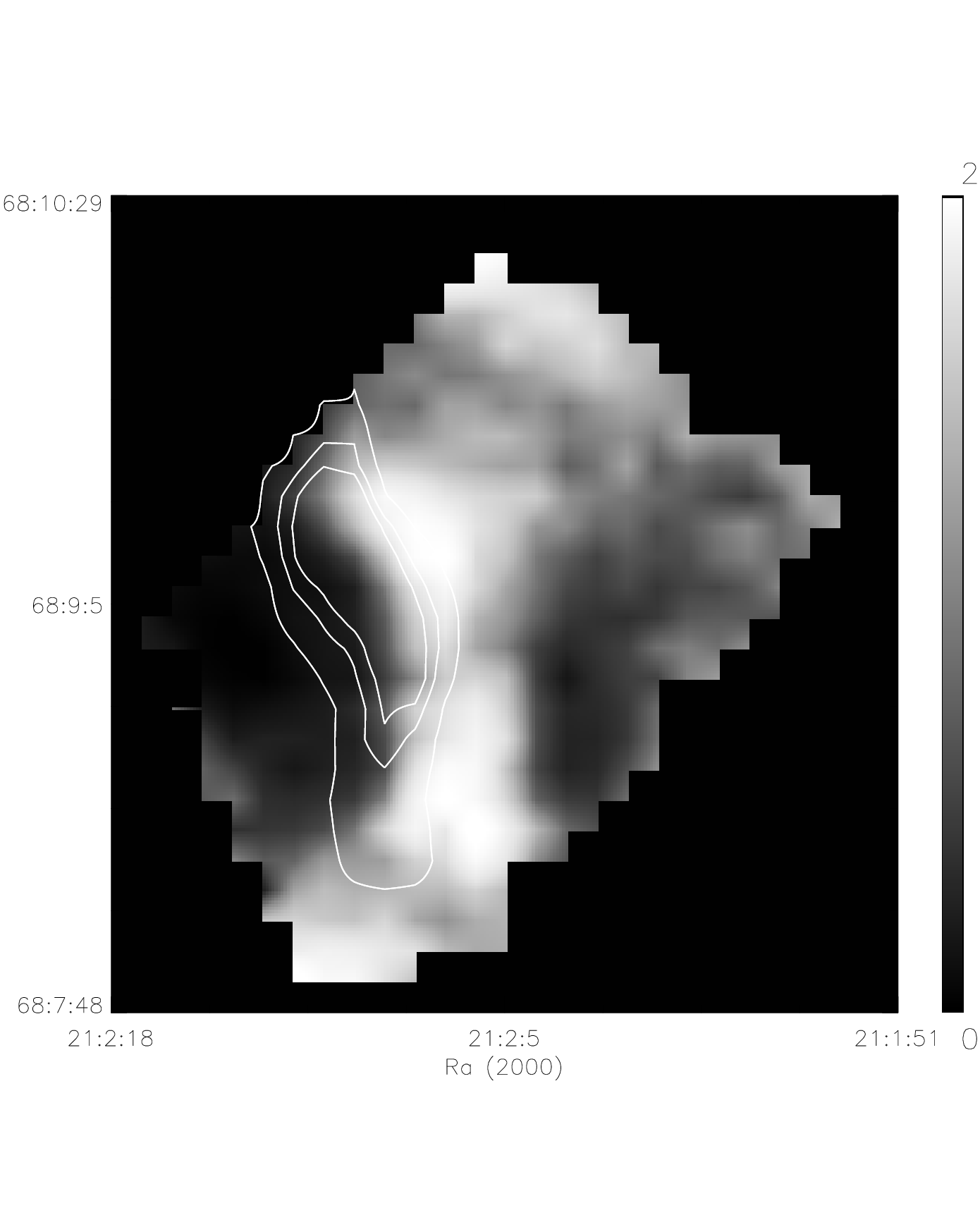}
\caption{ Map of the emission ratios \emph{Signal 1}/\emph{Signal 2} (upper) and  \emph{Signal 2}/\emph{Signal 1} (lower)
 in NGC 7023-E. The 5-35 $\mu$m integrated mid-IR emission of the filament is in contours
 (contours 1, 2, 3 are respectively 1, 1.2, 1.4 $\times$10$^{-4}$ W m$^{-2}$ sr$^{-1}$). 
The illuminating star lies West of the filament.}
\label{map7023}
\end{center}
\end{figure}

\begin{figure}
\begin{center}
\includegraphics[width=\hsize]{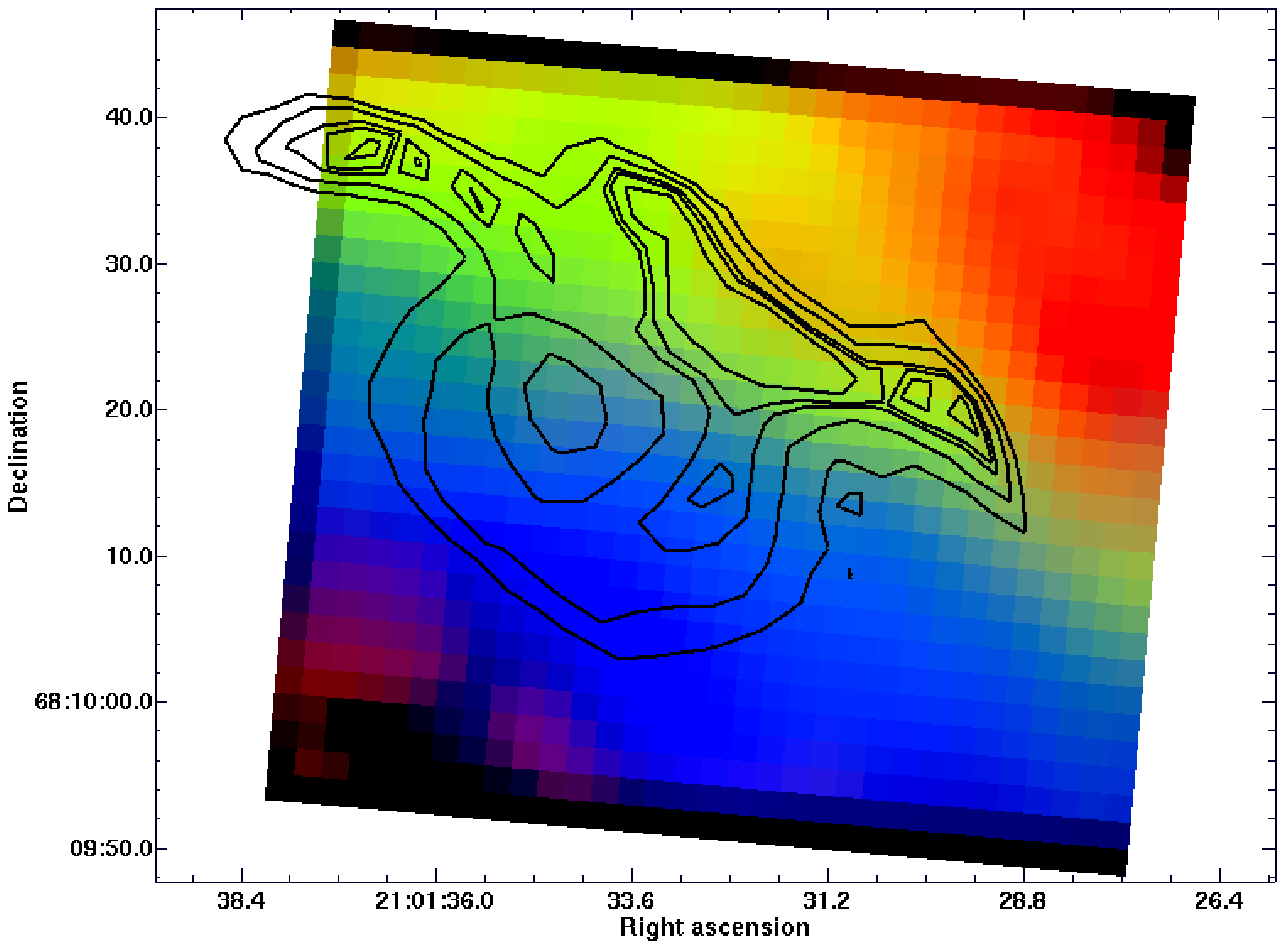}
\caption{Map of the correlation $c_p$ of each extracted signal with observations, 
\emph{Signal 1} in red, \emph{Signal 2} in green and \emph{Signal 3} in blue. The contours
are the Infrared Array Camera (IRAC) emission at 8 $\mu$m showing the filamentary
structure of this PDR. The illuminating star is situated in the lower left hand corner. 
In this region a slight artifact correlation of \emph{Signal 1} with observations is 
seen due to the presence of a continuum emanating from big grains at thermal equilibrium.}
\label{map7023NW}

\end{center}
\end{figure}

The observed data in matrix $X$.
may be expressed
as the product of two matrices, $W$ and $H$ eq. (5). 
Therefore, each observed spectrum from $X$ is defined as a linear combination of the extracted spectra in $H$.
In our case, each IRS spectrum at 
position
$(p_x, p_y)$ in a given cube can be written as follows:

\begin{equation}
\label{specr}
Obs(p_x, p_y, \lambda)= \sum_{n} w(p_x, p_y)_{n} {S_n(\lambda)} 
\end{equation}
where $Obs (p_x, p_y, \lambda)$ is the observed spectrum from the pixel indexed by $(p_x, p_y)$,
$S_n(\lambda)$ is the 
$n^{th}$
extracted spectrum 
in $H$ and $w(p_x, p_y)_{n}$ are the weights relative to spectrum $n$.
In our case $n$ is restricted to a maximum of 2 extracted spectra, except in the case of NGC 7023-NW
where $n=3$. 
The NMF algorithm 
extracts the sources up to arbirary scaling factors, i.e. it
provides $y_p=\eta_p S_p$, where $\eta_p$ is 
an unknown scale
factor and $S_p$ is the 
$p^{th}$
"source".
Each
output can be compared to the observations using the correlation 
between observations and NMF output.
We centered the observations 
(and therefore the outputs)
and considered that the extracted spectra are not correlated.
Thus the correlation 
parameter
$c_p$ we consider here can be written:
\begin{equation}
\label{corr}
c_p=E[Obs(p_x, p_y, \lambda) y_p]= \eta_p w(p_x, p_y)_{p} E[S_p(\lambda)^2] 
\end{equation}
where $E[.]$ stands for expectation.
We can then calculate $\gamma(p_x,p_y)=c_p/c_{p'}$ the ratio of the 
correlations
of Signal $p$ and Signal $p'$ to the observations:
\begin{eqnarray}
\label{map}
\gamma(p_x, p_y)&=&\frac{E[Obs(p_x, p_y, \lambda) y_p]}{E[Obs(p_x, p_y, \lambda) y_{p'}]} \nonumber\\
&=&\frac{ w(p_x, p_y)_{p}}{w(p_x, p_y)_{p'}} \times 
\underbrace{ \frac{E[S_p(\lambda)^2]  \eta_p}{ E[S_p'(\lambda)^2]\eta_{p'} }}_{\kappa}
\end{eqnarray}

where $\kappa$ is a constant for a given object. $\gamma$ traces the ratio of the weights $w(p_x, p_y)$ at the position $(p_x, p_y)$, and thus the emission ratio
between sources $p$ and $p'$, 
up to
the $\kappa$ 
scale
factor,
which does not depend on position $(p_x, p_y)$.
The maps of Figs.~\ref{map201}-\ref{map7023} show the spatial distribution
of the $\gamma$  emission ratio between \emph{Signal 1} and \emph{Signal 2},
\textbf{and Fig.~\ref{map7023NW} shows the $c_p$ correlation map for each signal}.

In Ced 201 the star is located in the central part of the nebula, and the \emph{Signal 2} emission 
is also concentrated in this region (Fig.~\ref{map201}), whereas \emph{Signal 1} dominates 
in the periphery of the nebula (Fig.~\ref{map201}).
In Oph-fil, the star is West of the filament. Again, \emph{Signal 2} is dominant in regions closer
to star (Fig.~\ref{mapOph-fil}) whereas \emph{Signal 1} emits in deeper regions
(Fig.~\ref{mapOph-fil}).
In NGC 7023-E, the illuminating star is located West of the filament. The distribution maps show that
the maximum \emph{Signal 2} emission is on the edge of the filament (Fig.~\ref{map7023}).
Further East in the cloud, the emission is highly dominated by \emph{Signal 1} (Fig.~\ref{map7023}).
Finally, in the case of NGC 7023-NW  \emph{Signal 1} is dominant behind the filament, \emph{Signal 2}
is dominant on the edge of the filament facing the star, and \emph{Signal 3} in front of the filament,
in the closest region to the star (Fig.~\ref{map7023NW}).


\section{Efficiency of the reconstruction of the observations using
the extracted spectra}\label{discussion}

BSS methods are quite new to astrophysics, and for this reason they should be used carefully.
However, several arguments confirm that in our particular case, NMF can be used efficiently.

\begin{figure}[h!]

\includegraphics[width=9cm]{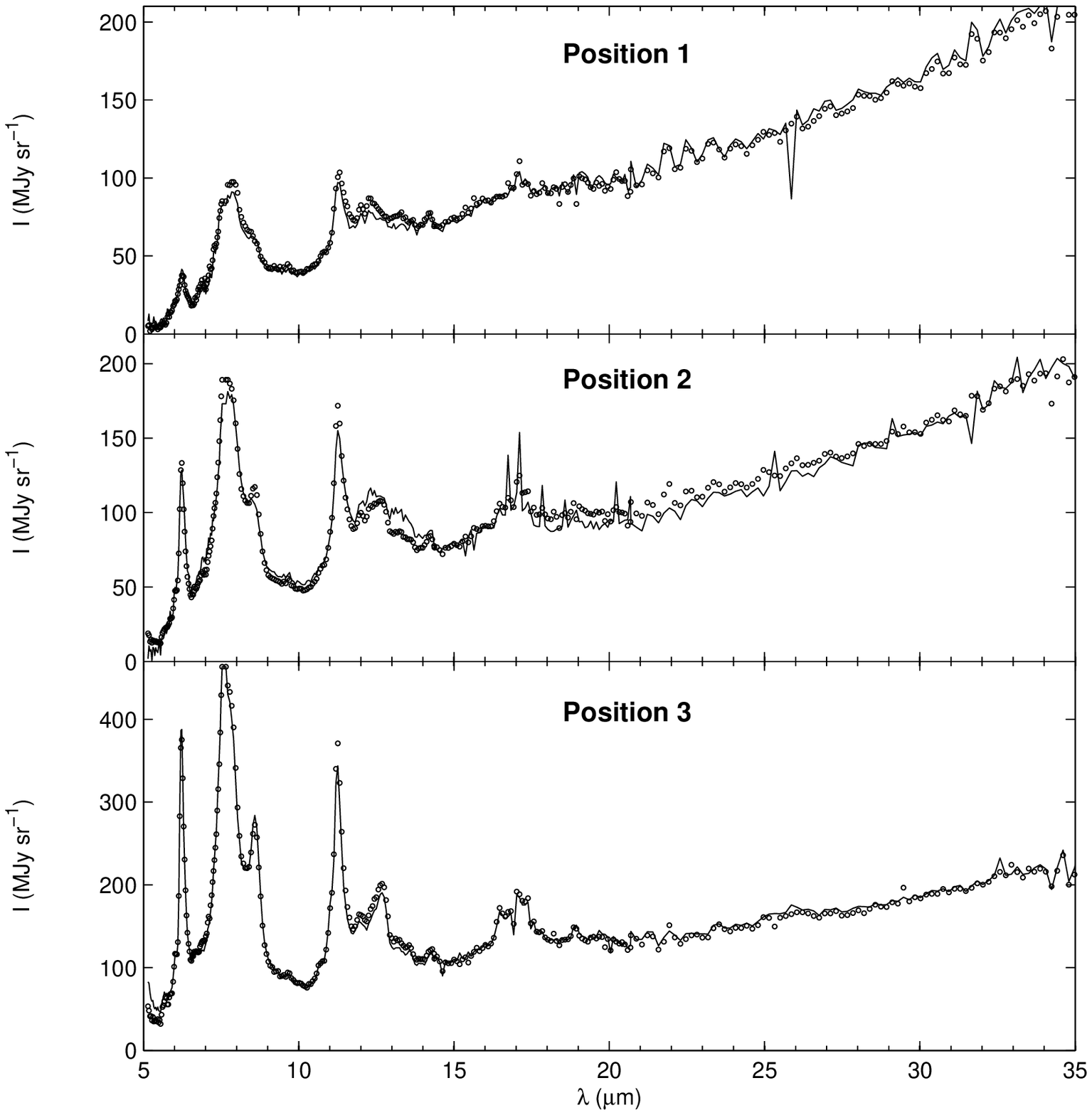}
\vspace{-0.5cm}
\caption{Reconstruction of the observations on three pixels of the Ced 201 cube,
 using linear combinations of the extracted spectra. Solid lines are the observations
and circles show the reconstruction.
Position 1 is at the periphery ~30" North of the central star, Position 2 is 8" North of the star and Position
3 is very close to the star. 
}

\label{recons}
\end{figure}

\begin{figure}[h!]

\includegraphics[width=9cm]{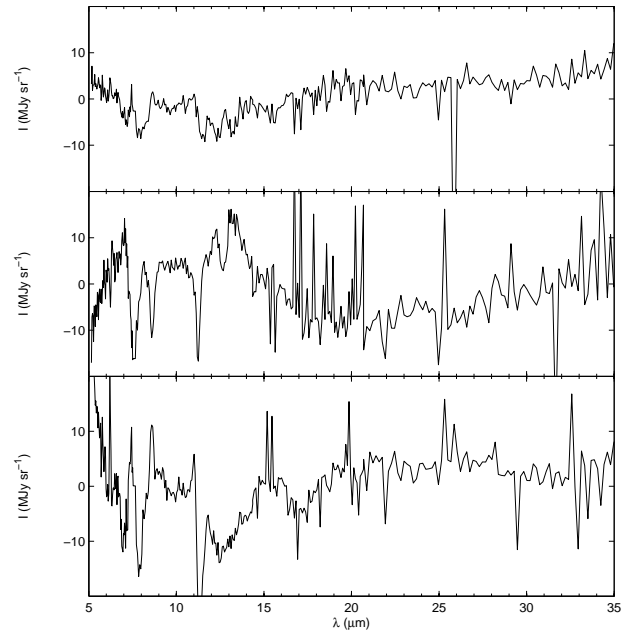}
\vspace{-0.5cm}
\caption{Residuals from the fit presented in Fig.~\ref{recons} (i.e. observation minus reconstruction).}

\label{residu}
\end{figure}

Contrary to most approaches,  there is absolutely no \emph{a priori} 
information included in the algorithm, and no subjective physical constraints 
on the source spectra, so that the extracted spectra are purely based on the information
contained in the data.
Thus, it is quite convincing that we are able to recover a spatial structure 
without any information on this included in the algorithm.
In this paper, we show that for the studied PDRs, the mid-IR emission spectrum can be fitted everywhere
using linear combinations of two or three simple spectra extracted from the considered PDR.
Figure~\ref{recons} shows the reconstruction of the observations taken on
three points of the Ced 201 PDR. 
The first position is at the periphery of the PDR, the second one is closer to the star and the last 
one is very close to the star. Using linear combinations of the extracted signals (see Sect.~\ref{Xtract}) 
we can reproduce the observations accurately, for the three positions. The residuals from this
fit are shown in Fig.~\ref{residu}. The ratio of the power of the observed signal to the power the 
of residuals is about 30 for position 2 (see Fig.~\ref{recons}), which is consistent
with the value of the SNR we have estimated (see Sect.~\ref{test}). The residuals are at noise level, proving
that the reconstruction from the extracted spectra of observations is efficient.
It is remarkable that we are able to show that in some regions of Ced 201, nearly all
the mid-IR emission is due to \emph{Signal 1}. Also note that, even though we chose
to use NMF for the reasons exposed in Sect.~\ref{choice}, 
\emph{FastICA} which is a completely different method provided similar results,
confirming their relevance.

\section{Nature of the carriers of the extracted spectra} \label{interpretation}


\begin{table*}[ht!]
\caption{Central wavelengths of the emission bands in the extracted spectra. The $\nu I_{\nu}$ integrated intensity ratios of the bands 
compared to the  7.7$\mu$m band are written in parentheses. The bands were fitted using lorentzian profiles. }
\label{table2}
\begin{center}
\begin{tabular}{cccccccc}

\hline \hline
&\multicolumn{3}{c}{VSGs}&&
\multicolumn{3}{c}{PAHs (neutral and cation mixture)} \\

\cline{2-4}
\cline{6-8}

\noalign{\smallskip}
Bands&Ced 201 & Oph-fil& 7023-E&&Ced 201& Oph-fil & 7023-E\\
\noalign{\smallskip}
\hline


6.2 $\mu$m   &  6.2     &   6.21   & 6.27       &    & 6.24   &  6.23      &  6.24         \\

   &  (0.06)    &   (0.27)  &  (0.41)        &    & (0.35)  &   (0.38)     &   (0.37)             \\
\noalign{\smallskip}

7.7 $\mu$m   & 7.87    &   7.77  & 7.79   &    & 7.66  &  7.72       &  7.70       \\

  & (1.0)   &    (1.0) &  (1.0)   &    & (1.0)  &   (1.0)   &  (1.0)      \\

\noalign{\smallskip}
8.6 $\mu$m   &  -    &    8.54   &    8.51      &    & 8.60        &  8.57       &   8.58          \\

  & -   &    (0.05) &  (0.05)   &    & (0.13)  &   (0.16)   &  (0.21)      \\

\noalign{\smallskip}
11.3 $\mu$m  & 11.41  &  11.40  & 11.38     &     &11.26 &  11.29      &  11.28         \\
  & (0.19) &  (0.13)  & (0.13)   &    & (0.13) &   (0.26)    &   (0.30)        \\

\noalign{\smallskip}  
12.7 $\mu$m  & -  &   -    &   -      &    &12.59 &  12.66       &  12.62        \\

  & -   &    - &  -   &    & (0.11)  &   (0.21)   &  (0.25)      \\

\noalign{\smallskip} 
Cont.     &  Yes &  Yes     &  Yes  &    &  No/weak  &   No/weak        &    No/weak         \\
\hline


\end{tabular}
\end{center}
\end{table*}

\begin{table*}[ht!]
\caption{ Same as Table 1 but for NGC 7023-NW. }
\label{table3}
\begin{center}
\begin{tabular}{cccccccc}

\hline \hline
&\multicolumn{3}{c}{This work}&&
\multicolumn{3}{c}{RJB} \\

\cline{2-4}
\cline{6-8}

\noalign{\smallskip}
Bands&VSGs&PAHs$^0$&PAHs$^+$&&VSGs&PAHs$^0$&PAHs$^+$\\
\noalign{\smallskip}
\hline
6.2 $\mu$m   &  6.24     &   6.22   & 6.24       &    & 6.29   &  6.27      &  6.28         \\

   &  (0.28)    &   (0.34)  &  (0.35)        &    & (0.35)  &   (0.47)     &   (0.40)             \\
\noalign{\smallskip}

7.7 $\mu$m   & 7.74    &   7.64  & 7.61   &    & 7.82 &  7.65       &  7.63       \\

  & (1.0)   &    (1.0) &  (1.0)   &    & (1.0)  &   (1.0)   &  (1.0)      \\

\noalign{\smallskip}
8.6 $\mu$m   &  -    &    8.55   &    8.57      &    & -        &  8.58       &   8.59          \\

  &    &    (0.29) &  (0.27)   &    &  &   (0.31)   &  (0.27)      \\

\noalign{\smallskip}
11.3 $\mu$m  & 11.36  &  11.25  & 11.18     &     &11.36 &  11.30      &  11.18         \\
  & (0.10) &  (0.37)  & (0.17)   &    & (0.28) &   (0.42)    &   (0.12)        \\

\noalign{\smallskip}  
12.7 $\mu$m  & -  &   12.66    &   12.67    &    & - &  12.76       &  12.63        \\

  &  &  (0.16) &  (0.08)  &    &   &   (0.17)   &  (0.08)      \\

\noalign{\smallskip} 
Cont.     &  Yes &    No/weak    &   No/weak   &    &  Yes  &   No/weak       &    No/weak         \\
\hline


\end{tabular}
\end{center}
\end{table*}

\subsection{Continuum-dominated spectrum}
\label{XVSG}

In all the extractions, \emph{Signal 1} exhibits a very clear association of continuum, and wide
AIB emission (Tables~\ref{table2}, \ref{table3}). This particularity was attributed to the emission
of carbonaceous VSGs by \citet{ces00} in Ced 201, from ISOCAM observations covering a 
wavelength range from 5 to 17 $\mu$m. The authors also proposed that these grains 
should be found everywhere in the interstellar medium. RJB have shown that using mathematical 
analysis, it is possible to identify these grains in other PDRs than Ced 201, from ISOCAM 
observations. They also mention that in the emission of this population, the position of the 
"7.7" $\mu$m band is shifted towards higher wavelengths which is also what we observe 
(Tables~\ref{table2}, \ref{table3}). When compared to spectra 2 and 3, the main characteristics
of the \emph{Signal 1} bands are widths typically  50 \% higher and a clear  red-shift of
 the 7.7 and 11.3\,$\mu$m bands. 
The 8.6\,$\mu$m band is weak or  absent and it is not possible to extract the 12.7 $\mu$m band from 
the 12-15 $\mu$m plateau. The 15-18 $\mu$m region is difficult to study in details because of the 
presence of a strong H$_2$ emission at 17 $\mu$m. 

The IRS-LL module enabled us to build spectral  cubes up to 35~$\mu$m
(note that the residual fringes observed in the 20-30 $\mu$m region are instrumental effects and are
not due to the algorithm or signatures of the dust). We have shown that a single spectrum, thus a 
single continuum, can reproduce the continuum emission all over the Ced 201 PDR (see 
Sect.~\ref{discussion}). This indicates that the shape of the continuum does not vary significantly
with distance to the star,  though its intensity varies. From this observation, we can conclude that 
these grains are most likely transiently heated. If this was not the case, their temperature, and 
therefore the continuum shape, should strongly change as a function of the distance from the 
star, as it happens for classical big grains (BGs). However we cannot completely rule out the presence 
of a fraction of grains at thermal equilibrium emitting in the 25-35 $\mu$m range. On this basis, 
and because of the strong similarity of our \emph{Signal 1} spectrum and the ISO VSG spectra 
in the 5 to 16~$\mu$m region, either observed \citep{ces00} or extracted mathematically (RJB), 
we propose to attribute \emph{Signal 1} in the 5-25 $\mu$m range to the emission of VSGs 
which appear to be carbonaceous, and probably corresponding to the VSGs used in the 
\citet{des90} model. 
In the following section we fit the emission of the Ced 201 PDR using the model of \citet{des90}
in order to provide some clues on the contribution of the two populations of grains (VSGs and BGs)
to the continuum.

\subsection{Modelling the emission of VSGs versus BGs}\label{model}

\begin{figure}[!h]
\includegraphics[width=\hsize]{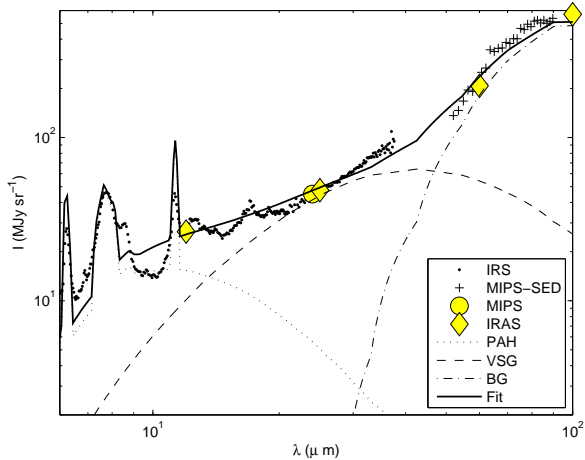}
\caption{Spectral energy distribution (5-100 $\mu$m) of the Ced 201 reflection nebula. The continuous line is the
emission modeled according to the dust model of \citet{des90} with the parameters from Table~\ref{des}.}
\label{fitall}
\end{figure}

The problem of calculating the infrared emission from VSGs in the ISM 
 has been addressed by several authors \citep{des90,sie92,dra01,dwe04}.
One question we would like to address is whether BGs contribute to the continuum
below 35 $\mu$m in the studied PDRs.
Although this point would deserve a full study by itself, we provide here first elements.
To estimate the emission of BGs, we fitted the total IR emission
from Ced 201 gathering the IRS, MIPS, MIPS-SED and IRAS data, using the model of 
\citet{des90}. The input parameters we have used for the model are listed in Table~\ref{des}.
The radiation field was calculated using the Meudon  PDR code \citep{leb93,lep06}.
The stellar spectrum is from the Kurucz library with a temperature of
 10{\,}500~K \citep{kur91} and a radius of 2.26\,$R_{\sun}$. The distance to the cloud, d,  was adjusted
to obtain the value of G$_0$ =300 at the interface as given by \citet{you02}. This gives d=0.0145pc.
This PDR has the strongest UV field of the three PDRs for which we have the full 5-35 $\mu$m data, and 
should thus host the hottest BGs, likely to emit in the mid-IR continuum.
The resulting fit (Fig.~\ref{fitall}) shows that the emission from VSGs is dominant up to 50 $\mu$m.
However, because this model is not completely adapted to the grain populations we consider, and because 
of the difficulties to calibrate the IRAS data, it is not possible to rule out BGs emission over 25 $\mu$m.
Moreover, as mentioned by \citet{dwe04}, the optical properties of BGs and VSGs are not well constrained,
implying that their size distribution cannot be derived precisely. 
Still, the model provides qualitative information on the emitting population, and
we can conclude from the fit that the emission in the 25 $\mu$m IRAS band is dominated by stochastically heated VSGs.

\begin{table}[ht!]
\caption{Input parameters for the fit of the Ced 201 IR emission with the model of \citet{des90}.}
\label{des}
\begin{tabular}{ccccc}
\hline \hline
\noalign{\smallskip}
 Component & Relative mass abundance& $\alpha^{*}$ & a$_{min}$ & $a_{max}$    \\
\hline
 PAHs & $3.4  10^{-4}$ & 3.0   & 0.4  & 1.2   \\

 VSGs & $5.7 10^{-3}$ & 2.6 & 1.2 &  10  \\

 BGs  & $0.994$ & 2.0 & 10 & 500   \\
\hline
\multicolumn{5}{l}{\tiny{* $\alpha$ is the  exponent in the power law of the size 
distribution: $n(a) \propto a^{-\alpha}$ where $a$ is the size in  }}\\
\vspace{-0.5cm}\\
\multicolumn{5}{l}{\tiny{ nanometers and $n(a)$ gives the numer density of grains with a radius between $a$ and $a + da$}}\\
\\

\end{tabular}
 \end{table}

\subsection{Pure band spectra: PAHs}
\label{pureband}

\emph{Signal 2} is dominated by the AIBs and is similar to the PAH-like emission of RJB. 
Therefore, we propose that \emph{Signal 2} is due to free PAH emission.
The three extracted PAH spectra of Figs.1-3, are very similar.
In this case, the 11-14 $\mu$m region is dominated by the 11.3 and 12.7 $\mu$m bands and no plateau
such as the one found in the VSG spectra is detected. Here also the 15-18 $\mu$m region is difficult to study because of
the high contamination by the H$_2$ line at 17 $\mu$m line.
However, the spectrum extracted in the case of Ced 201 exhibits a lower 11.3 $\mu$m band than in NGC 7023-E and
Oph-fil when compared to the intensity of the "7.7" $\mu$m band (see Table~\ref{table3}). This variation of 
the 11.3/7.7 $\mu$m ratio was interpreted by  \citet{job96,slo99,bre05} as an effect of variation in the relative
abundance of neutral and positively ionised PAHs  (respectively PAHs$^0$ and PAHs$^+$ hereafter), and \citet{fla06}
have shown that the effect of size on this ratio was weak.
Thus, the PAH extracted spectra in Ced 201, Oph-fil and NGC 7023-E are mixtures of neutral and cationic PAHs.
 In NGC 7023-NW, RJB were able to identify the 
emission spectra of two populations respectively dominated by PAHs$^0$ and PAHs$^+$.
In this object we were also able to extract two pure band spectra
similar to those determined by RJB which we also attribute to PAHs$^0$ and PAHs$^+$
(see Fig.\ref{ngc7023nw}).This strengthens the validity of both results.
Note however that the PAH$^+$ spectrum we extract here has a higher 11.3 to 7.7 $\mu$m 
integrated intensity ratio than the one
of RJB (see Tables \ref{table2}, \ref{table3}). This is because the observations we analyse here do not cover the regions close
to the star. Thus our PAH$^+$ population is "less" ionised than the one extracted by
RJB, which we will consider as reference hereafter. On the other hand, the PAH$^0$ we have extracted is
close to the one of RJB (see Table~\ref{table3} and Fig.~\ref{ngc7023nw}).
This extracted spectrum is also similar to the one observed in the
HII region of the Horsehead nebula by \citet{com06},
where PAHs are found to be purely neutral because of the extreme abundance of electrons.
From this we can conclude the following: 

\begin{itemize}
\item{the PAH$^+$ spectrum extracted by RJB is dominated by PAH cations. 
In the following, we consider this spectrum to be representative of cation emission. 
This was confirmed by \citet{fla06};}
\item{the PAH$^0$ spectrum can be attributed to a population containing only neutral PAHs.}
\end{itemize}

\subsection{Ionisation of PAHs in Ced 201, Oph-fil and NGC 7023-E }\label{ionisation}

As mentioned in Sect.~\ref{pureband}, the PAH spectra of Ced 201, Oph-fil and NGC 7023-E are mixtures 
of neutral and cation PAH emission. The reason why we were not able to extract the individual 
PAH$^0$ and PAH$^+$ spectra in these objects is that the spectral variations are not strong enough
in these cases to provide sufficient information to the algorithm to disentangle the two populations.
In order to estimate the normalised fraction of positively ionised PAHs $n_+/(n_0+n_+)$  in these PDRs,
we used the PAH$^0$ and PAH$^+$ spectra extracted from the data of NGC 7023-NW by RJB.
By comparing the $\nu I_{\nu}$ integrated intensity ratios between the 7.7 and 11.3 $\mu$m bands
of these extracted spectra (Table~\ref{table3}), to the pure band spectra of Ced 201,
NGC 7023-E and Oph-fil (Table~\ref{table2}), we could estimate $n_+/(n_0+n_+)$,
using NGC 7023-NW as a reference.
This ratio is equal to 0.97, 0.53, and 0.40 for Ced 201, Oph-fil and NGC 7023-E respectively (Table~\ref{table5}).
This result shows that Ced 201 has a much higher proportion of ionised PAHs, probably because of its 
low density (see Table~\ref{table4}) implying a low recombination rate with electrons. 
RJB and \citet{job96} also show that a large proportion of ionised PAHs is found in NGC 7023-NW and 
NGC 1333-SVS3 respectively. This is likely due to the fact that these PDRs are close to the illuminating star.
NGC 7023-E and Oph-fil, are dominated by neutral PAHs, probably because these PDRs are far from the star
and have a higher density with respect to Ced 201 (see Table~\ref{table4}).


\citet{fla06} have defined the R$_{7.7/11.3}$ parameter as the ratio between the $\lambda I_{\nu}$ integrated intensity
of the 7.7 and 11.3 $\mu$m band. We have calculated this parameter for Ced 201, NGC 7023-E and Oph-fil (Table~\ref{table5}).
\citet{fla06} have shown that the R$_{7.7/11.3}$ parameter is related to the ionisation parameter 
$G_0\sqrt{T}/n_e$, where $G_0$ is the UV field in units of the Habing field,
$T$ is the gas temperature and $n_e$ the electronic density.
Using this result, we calculated the ionisation parameter from the  PAH spectra of our PDRs 
with the R$_{7.7/11.3}$ ratio from Table~\ref{table5}. 
We also calculated this parameter independently using the physical parameters  from Table~\ref{table4}.
For this calculation, we used a $n_e/n_H$ ratio of $ 1.4\,10^{-4}$ \citep{sno95}.
The results of these two calculations are reported in Table~\ref{table5}.
The estimates of the ionisation parameter found here are consistent with the ones calculated from the physical parameters.
This shows that our method to quantify the ionisation state of PAHs is consistent with the ones 
of \citet{fla06} and could be applied to any interstellar PAH spectrum.


\begin{table*}[ht!]
\caption{Physical conditions of the PDRs.}
\label{table4}
\begin{center}
\begin{tabular}{cccccc}

\hline \hline
\noalign{\smallskip}
 Object     & Star: ST, T(K) & n$_H$ (cm$^{-3}$) & T(K) & G$_0$ (Habing) & Ref. \\

\hline
\noalign{\smallskip}
Ced 201     & B9.5V, 10 400  & $4.10^2$          & 200 & 300 & \citet{you02} \\
\noalign{\smallskip}
Oph-fil     & B2V, 22 000    & $3.10^4$          &  310 & 100 & \citet{hab03} \\
\noalign{\smallskip}
NGC 7023-E  & B5e, 15 000    &  $7.10^3$        &  340 & 87    & \citet{rap06}  \\
\hline
\\

\end{tabular}
\end{center} 
\end{table*}

\begin{table*}[ht!]
\caption{Estimated parameters in the studied PDRs: the R$_{{7.7}/{11.3}}$  band intensity ratio, the ionisation
parameter, the UV field and the ionisation fraction.}
\label{table5}
\begin{center}
\begin{tabular}{ccccc}

\hline \hline
\noalign{\smallskip}
 Object      & R$_{{7.7}/{11.3}} $ & G$_0~\sqrt{T}/n_e$ & $G_0$ &   $n_+/(n_0+n_+)$ \\

\hline
\noalign{\smallskip}
Ced 201  &   3.76 & 105000$^1$, 75000$^2$   & 300     & 0.97$^3$ \\
\noalign{\smallskip}
Oph-fil   & 1.89   &   1060$^1$,420$^2$       & 100     & 0.53$^3$\\
\noalign{\smallskip}
NGC 7023-E   &  1.63 &    1040$^1$,1650$^2$   & 87     & 0.40$^3$  \\
\hline
\multicolumn{2}{l}{\tiny{$^1$Using the results from the model described in \citet{fla06}}}
&\multicolumn{2}{l}{\tiny{$^2$Using the data from Table~\ref{table4}}}\\
\vspace{-0.5cm}\\
\multicolumn{2}{l}{\tiny{$^3$Using the method described in this paper}}
&\multicolumn{2}{l}{\tiny{}}
\vspace{-0.1cm}\\
\\

\end{tabular}
\end{center} 
\end{table*}

\section{Chemical evolution of the very small dust particles in the observed PDRs}

The distribution maps presented in Sect.~\ref{distribution} clearly show that VSGs and PAHs$^0$/PAHs$^+$
 are not situated in the same regions of the studied PDRs. 
Going towards the star in the PDRs, VSGs followed by PAHs and
eventually PAHs$^+$ (case of NGC 7023-NW) are successively dominant. 
This evolution of the dust populations across the PDRs while the UV flux is becoming greater
 suggests that these three populations are chemically linked.
The disappearance of VSG emission, while the PAH emission increases, 
suggests that there is transformation of VSGs into 
PAHs  under the action of the UV field, as proposed by \citet{ces00} for Ced 201 and by RJB 
for $\rho$Oph-SR3 and NGC 7023-NW PDRs.
Here, we report new evidence of this transformation  in three regions: Ced 201, NGC 7023-E and Oph-fil PDRs.
This is consistent with a scenario in which VSGs could be PAH clusters as suggested by RJB.
Interestingly, recent laboratory experiments  have shown that coronene clusters can be photo-evaporated into free 
coronene units \citep{brec05}.
PAH aggregates would likely form in cold dense regions \citep{bou90,ber93,rap06}, and their photo-chemistry 
would start when the UV field is strong enough. 
Thus, the observed variations of the mid-IR interstellar spectra
\citep{pee02, wer04, bre05}  have to be explained as the evolution of a mixture of PAHs and VSGs
under the effect of UV flux.
The processing of VSGs can explain the diminution of continuum emission while
approaching the irradiating star reported by  \citet{wer04}. \citet{bre05} have found that the 7.7 $\mu$m
emission band shifts towards shorter wavelengths as the UV field increases, which they attributed to the evolution from anionic
to cationic species.
According to our scenario, this shift can be explained as the chemical evolution of the mixture, from
VSGs, with a band between 7.74 and 7.87 $\mu$m, into PAHs with a band between 7.63 and 7.70 $\mu$m.
In that previous work, the authors show that the R$_{7.7/11.3}$ ratio is increasing when moving away from
the exciting source, tracing the ionisation state. However, after a certain distance, the ratio reaches its 
maximum and starts decreasing again. This was explained by the authors as a result of a greater proportion
of anionic PAHs, but is more likely consistent with the presence of VSGs which exhibit a low R$_{7.7/11.3}$ band ratio 
(see Tables~\ref{table2},\ref{table3}).

\section{Conclusion}

Using the data from the Infrared Spectrograph onboard Spitzer,
combined with powerful Blind Signal Separation methods, we
were able to extract the spectra of two types of very small interstellar
dust particles: one carrying mainly the AIB features and the other
a mid-IR continuum and broad AIBs. These two populations are identified
as PAHs and VSGs respectively.  Concerning the pure AIB spectra,
we could identify in the case of NGC 7023-NW the emission from PAH$^0$
and PAH$^+$ dominated populations. Using the spectra of RJB,
we were able to simply estimate the ionisation fraction of PAHs
in various PDRs.
Cesarsky et al. (2000) proposed that VSGs should
be found everywhere in the interstellar medium. However their
detection is far from being obvious. Here we have shown that
BSS methods enable us to extract their spectrum taking advantage
of both the spatial and spectral information available
in the IRS spectral cubes. Thanks to the wide range of wavelengths
covered, it was possible to confirm that these grains are responsible
for the interstellar mid-IR continuum and could dominate the emission
up to 50 $\mu$m in cool PDRs. The similarities of their spectral  
features with the AIBs show that they are carbonaceous.  This is consistent  
with the predictions of \citet{des86} that VSGs are mostly graphitic and
not silicates. The distribution maps of PAHs and VSGs further support  
this conclusion. We indeed show that there is a transformation of VSGs
into PAHs, probably under the effect of the UV flux.
PAHs and VSGs clearly dominate the emission in the 12 and 25 $\mu$m
IRAS bands respectively.
This result is consistent with the work of \citet{fue92} who
proposed that the 25 $\mu$m emission in reflection nebulae was
due to another type of grains than the ones emitting at 12 $\mu$m.
It is also consistent with the results of \citet{ver00}, who
have shown that the cloud-to-cloud variations of the mid-IR
emission of molecular cirrus can only be explained by changes
in the relative abundance of PAHs and VSGs.
The evolution from VSGs to PAH molecules can explain the commonly
observed mid-IR spectral variations.
The recently available data from the MIPS photometer in the 
Spectral Energy Distribution mode (MIPS-SED) is currently under 
analysis in order to learn more about the properties of VSGs at longer wavelengths.
The PAH$^0$, PAH$^+$ and VSGs spectra extracted here can be used as a basis to probe
the composition of the very small particles in various environments, including external galaxies.
To a certain extend, this could provide information on the local physical conditions.

\begin{acknowledgements}

The authors wish to acknowledge all the members of the SPECPDR team for their
contribution to the sucess of the proposal.  

\end{acknowledgements}

\bibliographystyle{aa}
\bibliography{biblio}

\end{document}